
\magnification=1200
\hfuzz=10pt
\hsize=4.8in
\vsize=7.3in
\baselineskip=16pt
\hoffset=0.35in
\voffset=0.1in
\parskip=.2cm
\parindent=2.5em
\tolerance=1000
%
\def\newline{\hfil\break}
\def\newpage{\vfil\eject}

\def\title#1{
{\baselineskip 14pt \tolerance=10000
\vglue 2truecm
\bf \noindent {#1}} \par}

\def\author#1{ \vskip 12pt \tolerance=10000
\noindent {\it #1} \par}
\def\adress#1 {\vskip 10pt \tolerance=10000
\noindent {#1} \par}
\def\adre#1 {\vskip 0pt \tolerance=10000
\noindent {#1} \par}
\def\abstract#1{
\vskip 30pt
{\noindent {\bf Abstract.\ }#1 }}
\countdef \sectno=11
\sectno = 0
\countdef \sbsectno=12
\sbsectno=0
\def\section#1 {\advance \sectno by 1 \sbsectno = 0
\vskip 24pt
{\tolerance=10000
\noindent {\bf \number \sectno. \ {#1}}}
\vskip 12pt \par}
\def\subsection#1 {\advance \sbsectno by 1
\if num \count12>1 \vskip 24pt \else \fi
{\tolerance=10000
\noindent {\bf \number \sectno. \number \sbsectno {#1}}}
\vskip 12pt \par}
\countdef \figno=13
\figno=0
\def\fig#1#2{\advance \figno by 1 \tolerance=10000
{\topinsert \vskip#1
\smallskip
\noindent {Fig.~\number \figno: {#2}}
\vskip 24pt \endinsert }}
\def\fih#1#2{\advance \figno by 1
{\hsize 8.5truecm \tolerance=10000
\topinsert \vskip#1
\noindent {Fig.~\number \figno:\ {#2}}
\vskip 24pt \endinsert }}

\title{LINEAR AND NONLINEAR MECHANISMS OF INFORMATION PROPAGATION}
\author{Antonio Politi$^{\ (1)}$ and Alessandro Torcini$^{\ (2)}$}
\vskip .2cm
\item{(1) -} Istituto Nazionale di Ottica, I-50125 Firenze, Italy;
INFN - Sezione di Firenze.
\item{(2) -} Dipartimento di Fisica, Universit\`a di Firenze;
INFM and GNSM - Sezione di Firenze; I.T.G. "A. Gramsci", I-50047 Prato,
Italy.

\abstract{The mechanisms of information transmission are investigated
in a lattice of coupled continuous maps, by analyzing the propagation
of both finite and infinitesimal disturbances. Two distinct regimes are
detected: in the former case, both classes of perturbations spread
with the same velocity; in the latter case, finite perturbations propagate
faster than infinitesimal ones. The transition between the two phases is
also investigated by determining the scaling behaviour of the order
parameter.}

\newpage
\noindent
Recently, it has been discovered that coupled-map lattices (CML) can exhibit
irregular behaviour even if the maximum Lyapunov exponent is negative
[1, 2]. This phenomenon has been called ``stable chaos'' [2]
in order to stress the apparent incongruity. As a matter of fact, a periodic
configuration is always approached in a finite time, on a lattice of length
$L$.
However, the average transient time $T(L)$ diverges exponentially with $L$.
In the thermodynamic limit $L \to \infty$, the statistical behaviour
exhibited for
times shorter than $T(L)$ appears to be stationary. Therefore, the usual
chaotic
indicators (as, e.g., Lyapunov exponents) can be meaningfully applied to the
characterization of the transient evolution as well.
In analogy with standard transient chaos, the value of the Lyapunov exponents
depends on the order the limits $L \to \infty$ and $t \to \infty$ are taken.

The reference model considered in this Letter is a coupled-map lattice based on
the usual diffusive-coupling scheme [3],
$$
   x_{t+1}^i =  f(\bar x_t^i) \eqno (1a)
$$
$$
  \bar x_t^i = {\varepsilon \over 2} x^{i-1}_t +
                (1 - \varepsilon) x^i_t +
                {\varepsilon \over 2} x^{i+1}_t  \eqno (1b)
$$
where $t$ is the discrete time, $i$ labels the lattice sites, $\varepsilon$
measures the coupling strength and periodic boundary conditions
are assumed ($x_t^{0} = x_t^L$, $x_t^{L+1} = x_t^1$).

In both [1] and [2], the map $f$ responsible for stable chaos
is of the type
$$
   f(x)  = \cases{ p_1 x + q_1  & $0 \le x \le x_c$ \cr
	 1 - (1-q_2)(x-x_c) / \eta & $x_c < x \le x_c  + \eta$ \cr
       q_2 + p_2(x-x_c-\eta) & $ x_c +\eta < x \le 1$}\quad , \eqno(2)
$$
($x_c = (1-q_1)/p_1$) for $\eta=0$, i.e. it is discontinuous and
piecewise linear.
The single map considered in Ref. [1] ($f_1$), corresponding to the parameter
values $p_1 = p_2 = 0.91$, $q_1=.1$, and $q_2=0$, is characterized by a
globally stable period-25 orbit. The parameters of the map in Ref. [2] ($f_2$)
are $p_1 = 2.7$, $p_2 = 0.1$, $q_1 = 0$ and $q_2 = 0.07$. In such a case, a
globally stable period-3 orbit exists.

One can easily conjecture that the sudden amplification of the distance
between points lying across the discontinuity (a phenomenon which escapes the
linear stability analysis) is at the origin of the random evolution. In fact,
Ershov and Potapov [4] have shown that nonchaotic turbulence persists in the
CML if the map $f_1$ is made continuous by introducing a finite $\eta
 < 4\cdot10^{-4}$. However, they have also shown that the continuization
of the map destabilizes the period-25 orbit, leading to a standard chaotic
attractor in the single map for $\eta > 5 \cdot10^{-3}$ Therefore, they
have concluded that the observed irregular behaviour is a reminiscence
of the usual transient chaos, characterized by a positive Lyapunov exponent.

We have found that, in analogy to what happens for the model of Ref. [1], the
dynamics of the CML with a continuous $f_2$ exhibits
a negative maximum Lyapunov exponent if $\eta < \eta _1 = 4 \cdot 10^{-4}$.
However, in our case, the critical value of $\eta$ corresponding to the
destabilization of
the period-3 orbit is much higher ($\eta_2 = p_1^2q_2 -1/p_1 \simeq 0.14$).
Therefore, the phenomenon of stable chaos has a less obvious
origin than conjectured in Ref. [4]. Nevertheless, one must conclude that
in both models, the linear stability analysis is not able to capture the
essence of the observed irregular behaviour.

We conjecture that two different mechanisms of information flow on the lattice
must be invoked: one based on the usual (local) linear instability, the other
based on ``nonlinear'' propagation phenomena, analogous to those present in
cellular automata [5]. The aim of the present Letter is to perform a
comparative study of the two mechanisms.

The CML has been analyzed for $\varepsilon = 2/3$. We investigate the evolution
of an initially localized disturbance $w_{t=0}^i = \delta_{i,0}$ in the
linear approximation. Its time evolution is of the type
$$
    w_t^i \simeq {\rm e}^{\Lambda(i/t)t} \quad ,     \eqno(3)
$$
where the comoving Lyapunov exponent $\Lambda(v)$ [6] is the growth rate
of the disturbance in a reference frame moving with velocity $v=i/t$. Within
the light-cone defined by
$$
\Lambda(v_{\ell}) = 0 \quad ,
$$
the perturbation is exponentially amplified. Accordingly, $v_{\ell}$ represents
the limit velocity of propagation of disturbances in the linear approximation.
The direct estimate of $\Lambda(v)$ may be affected by two technical problems:
({\it i}) a site-dependent normalization of the amplitude $|w_t^i|$ to cope
with
the different growth rates along the different lines $i = vt$; ({\it ii}) the
increasing number of sites that must be updated. These problems can lead to
serious difficulties, whenever the convergence towards the asymptotic shape is
slow. Deissler and Kaneko [6] applied the usual algorithm for Lyapunov
exponents
in a moving reference frame. This approach is not affected by the above
problems and it works perfectly in open-flow systems. However, in closed
systems
(like those ones we are currently investigating) boundary conditions strongly
affect the evolution of Lyapunov vectors.

This further problem can be overcome by following the approach recently
developed in [7]. The authors have introduced the specific Lyapunov exponent
$\lambda(\mu)$, defined as the maximum growth rate of a perturbation $w_t^i$
characterized by an
exponential profile, $w_t^i = \exp{(-\mu i)}\cdot u_t^i$. The new variable
$u_t^i$, assumed to satisfy periodic boundary conditions, obeys the evolution
equation
$$
u^i_{t+1} = f^\prime(y_t^i) [{\varepsilon \over 2} {\rm e}^{-\mu}
u^{i-1}_t + (1 - \varepsilon) u^i_t + {\varepsilon \over 2}
{\rm e}^{\mu} u^{i+1}_t]  \quad .
\eqno (4)
$$
The specific exponent $\lambda(\mu)$ can be estimated by applying the standard
algorithm developed for the computation of Lyapunov spectra.
The comoving exponent is related to $\lambda(\mu)$ through the Legendre
trasform [7]
$$
  \Lambda(v) = \lambda(\mu) - \mu \lambda^\prime(\mu) \quad  ;
  \quad  v  = - \lambda^\prime(\mu)
  \eqno(5)
$$
{}From the above equation,
$$
  v_{\ell} = - \lambda^\prime (\overline \mu) \quad ,
$$
where $\overline \mu$ is the $\mu$-value such that the first of Eq.~(5)
vanishes.
The derivative $\lambda^\prime(\mu)$ can be estimated with a good accuracy
by iterating a suitable recursive equation. After deriving Eq.~(4) with
respect to the parameter $\mu$ and introducing $z_t^i = (u_t^i)^\prime$, we
obtain
$$
z^i_{t+1} = f^\prime(y_t^i) [-{\varepsilon \over 2} {\rm e}^{-\mu}
u^{i-1}_t + {\varepsilon \over 2} {\rm e}^{\mu} u^{i+1}_t]
+ f^\prime(y_t^i) [{\varepsilon \over 2} {\rm e}^{-\mu}
z^{i-1}_t + (1 - \varepsilon) z^i_t + {\varepsilon \over 2}
{\rm e}^{\mu} z^{i+1}_t]
\eqno (6)
$$
The derivative of the specific exponent is then given by
$$
\lambda' (\mu)  =
{\sum_k z^k_{t+1} \cdot u^k_t \over
(t+1) (\sum_k u^k_{t+1} \cdot u^k_{t+1})^{1/2}}
\eqno (7)
$$
The results of simulations performed for several $\eta$ values are reported
in Fig. 1 (dashed line). The linear propagation develops for $\eta > \eta_1$,
when the maximum Lyapunov exponent becomes larger than zero.

The evolution of finite disturbances and the related nonlinear mechanisms can
be investigated only in terms of a direct approach. Let us consider two initial
configurations $\{ x^i_t\}$, $\{y^i_t\}$ differing by order ${\cal O}(1)$ in
a finite interval and coinciding outside. The corresponding configurations,
generated
after $T$ time steps, are then compared to determine the left and right borders
$i_l$, resp. $i_r$ of
the region where they differ more than a preassigned threshold $\theta$.
The propagation velocity $v_{n \ell}$ of finite disturbances is then defined as
$$
  v_{ n\ell} = \lim_{T \to \infty} {(i_r-i_l) \over 2T}  \quad .
$$
Numerical simulations reveal a sizeable dependence of the actual value of
$v_{ n \ell}$ on the computational accuracy. For instance, upon
increasing the number of significant digits from 8 to 64, the value of
$v_{ n\ell}$ increases by about 2 \% for $\eta = 0.01$.
This inaccuracy prevents a quantitative study of the scaling behaviour of the
difference $\Delta v = v_{ \ell} - v_{ n \ell}$ for
$v_{ \ell} \to v_{ n \ell}$. Such a difficulty has been circumvented by
extrapolating the numerically exact value of $v_{ n\ell}$ from the results
of increasingly accurate simulations [8].
As a matter of fact, the finite-precision estimates of the velocity converge as
$1/N_b^2$, where $N_b$ is the number of significant digits. The asymptotic
values of $v_{ n\ell}$ are reported in Fig. 1 (solid line).
Moreover, upon varying the value of the threshold $\theta$ from 1.$10^{-4}$ to
0.1, the resulting variation of $v_{ n \ell}$ is practically negligible.

Any infinitesimal perturbation $w^i_t$ is exponentially amplified within the
light-cone defined by $v_{\ell}$. Therefore, $w^i_t$ eventually becomes larger
than
any finite threshold $\theta$, so that the nonlinear velocity $v_{ n\ell}$
cannot be smaller than $v_{\ell}$. This inequality is indeed satisfied in the
simulations reported in Fig. 1, where two different ``phases'' are identified:
for $\eta \ge \eta_c = 0.013$, $v_{\ell} = v_{ n\ell}$ (I);
for $\eta < \eta_c$, $v_{ n\ell} > v_{\ell}$ (II). The former phase
corresponds to the usual regime found in lattices of logistic and cubic
maps [9].
The latter phase corresponds to a new regime, where the instability
due to a positive Lyapunov exponent is not sufficient to account for the
propagation of information along the chain. The examples analysed in Refs.
[1,2]
represent a limit case of this scenario, where there is no linear instability
at
all.

The transition between the two regimes can be investigated from the
dependence of the order parameter
$\Delta v = v_{\ell} - v_{ n\ell}$ versus $\eta_c - \eta$.
The numerical results reported in Fig.~2, reveal a scaling behaviour of the
type
$$
\Delta v = (\eta_c - \eta)^\gamma  \quad .
\eqno (8)
$$
The optimal scaling behaviour is obtained for the above reported value of
$\eta_c = 0.013$ yielding $\gamma \simeq 2$. Having we failed to derive even a
heuristic
explanation of this exponent, we have tried to reach a more complete
understanding of this phemomenon by studying the fluctuations of the effective
maximal comoving
Lyapunov exponent $\tilde \Lambda$. The effective exponent $\tilde \Lambda$
corresponding to a velocity $v$ over a time $t$ is defined as
$$
\tilde \Lambda \equiv { 1 \over t} \log \left \vert { y^i_t -
x^i_t \over \delta } \right \vert \quad ,   \eqno (9)
$$
where $i=vt$ and $\{x^i_t\}$, $\{y^i_t\}$ represent two
configurations which, at time $t=0$, differ by $\delta_0^0 = \delta$
in the site $i=0$, while coincide elsewhere.  Moreover, let
$P(\tilde \Lambda,v,t)$ denote the probability density to find $\tilde \Lambda$
in the interval $(\tilde \Lambda, \tilde \Lambda + d\tilde \Lambda)$.
In the limit $\delta \to 0$ and $t \to \infty$, the whole information about
the fluctuations of the Lyapunov exponent is conveyed by the multifractal
spectrum
$$
S(\tilde \Lambda,v) = { \log(P(\tilde \Lambda,v,t)) \over  t } \quad .
\eqno (10)
$$
The Lyapunov analysis has been performed for $\eta = 2 \cdot 10^{-3}$. For
this parameter value, linear and nonlinear velocities are $v_{\ell}=0.4184$,
$v_{n\ell}=0.5805$, respectively. We have computed the spectrum $S$ for
different velocities and time lags. The curves reported in Fig. 3 have been
all obtained for $t=40$; labels 1, 2 and 3 refer to
$\delta =10^{-7}$, $10^{-4}$ and $10^{-2}$, respectively. Finally, the spectra
reported in Fig. 3a correspond to $v=0.7$, while those ones in Fig. 3b to
$v = 0.5$.

If the amplitude of the perturbation is initially smaller than $\delta_c =
\exp [-\tilde \Lambda_{max}(v) t]$ (where $\tilde \Lambda_{max}(v)$ is the
maximum value assumed by the effective Lyapunov exponent $\tilde \Lambda$ at
velocity $v$), then it will always obey a linear equation during its
evolution over the first (arbitrary) $t$ time steps. In this case, one is
simply estimating the multifractal spectrum of the comoving Lyapunov exponent
$\Lambda$. On very general grounds, a bell shaped spectrum is
expected. This is in fact what observed for the curves labelled with ``1''
in Fig. 3.
If $\delta$ is increased above $\delta_c$, some perturbations become of
order ${\cal O}(1)$ and are then controlled by the full nonlinear equations.
One could simply conjecture that all $\tilde \Lambda$ values greater than
$ \tilde \Lambda_u =   -\log \delta /t$ are no longer detectable, while the
rest of the spectrum remains unchanged. This is precisely what happens when
the velocity $v$ is larger than $v_{n \ell}$ as, for instance, in Fig.~3a.
If instead, $v < v_{n \ell}$, a second peak appears at high $\Lambda$-values,
when $\delta$ is increased (see Fig. 3b). For $\delta$ sufficiently ``large'',
the new peak becomes
higher than the old one, meaning that the nonlinear mechanism prevails onto
the linear one.

The order of magnitude of the statistical error is much less than the amplitude
of the oscillations observed at small $\tilde \Lambda$s: they are intrinsic
fluctuations which are seen to decay slowly for increasing $t$. However,
this finite-size phenomenon does not prevent a clear-cut distinction between
the qualitative behaviour of the spectra observed in Fig. 3a and 3b.
Therefore, we can conclude that the exchange of
the two limits $\delta \to 0$ and $t \to \infty$ plays a crucial role in
the phenomenon studied in this Letter. The usual definition of Lyapunov
exponent is recovered if the former limit is taken first, while the
nonlinear propagation of information is revealed only by taking first the
latter limit.

In this Letter we have discussed two different mechanisms for information
propagation. We have identified a regime where the prevailing mechanisms
depends
on the amplitude of the propagating perturbation. Whenever nonlinear
processes eventually overtake the linear amplification, we expect different
statistical properties of the associated spatio-temporal pattern.
Finally, notice that the relevance of nonlinear mechanisms has been
pointed out also in a standard CML of logistic maps, with reference to
the predictability problem [10].

\newpage
\noindent
{\bf References}\par

\vskip .2cm
\item{[1]} J. Crutchfield and K. Kaneko, Phys. Rev. Lett. {\bf 60}, 2715 (1988)
\vskip .2cm
\item{[2]} A. Politi, R. Livi, G.-L. Oppo and R. Kapral, Europhys. Lett.
{\bf 22}, 571 (1993)
\vskip .2cm
\item{[3]} K. Kaneko, Prog. Theor. Phys., {\bf 72}, 980 (1984); {\bf 74}, 1033
(1985). I. Waller and R. Kapral, Phys. Rev. A, {\bf 30}, 2047 (1984).
\vskip .2cm
\item{[4]} S.V.Ershov and A.B. Potapov, Phys. Lett. A {\bf 167}, 60 (1992)
\vskip .2cm
\item{[5]} For a review , see {\it Theory and Applications of
Cellular Automata}, ed. S. Wolfram (World Scientific, Singapore, 1986).
\vskip .2cm
\item{[6]}  R.J. Deissler and K. Kaneko, Phys. Lett. A {\bf 119}, 397
(1987).
\vskip .2cm
\item{[7]} A. Politi and A. Torcini, Chaos {\bf 2}, 293 (1992).
\vskip .2cm
\item{[8]} An {\it ad hoc} routine has been developed to go beyond
the standard computer precision.
\vskip .2cm
\item{[9]} A. Politi and A. Torcini, unpublished.
\vskip .2cm
\item{[10]} G. Paladin and A. Vulpiani, J. Phys. A {\bf 27}, 4911
(1994).

\newpage
\noindent
{\bf Figure Captions}\par

\vskip .2cm
\item{Fig. 1} Linear (dashed) and nonlinear (solid) velocities
versus $\eta$. For the CML model (2) with $p_1 = 2.7$,$p_2 = 0.1$,
$q_1  = 0$ , $q_2 = 0.07$ and diffusive coupling $\varepsilon = 2/3$.
The two arrows indicate the position of the parameters $\eta_1 = 4 \cdot
10^{-4}$ and $\eta_c = 0.013$.

\vskip .2cm
\item{Fig. 2}  $\Delta v = v_{\ell} - v_{n \ell}$ versus $\eta_c - \eta$
($\eta_c = 0.013$) reported in a $\log$ - $\log$ scale. The circles represent
the actual point where the difference $\Delta v$ have been calculated.

\vskip .2cm
\item{Fig. 3}  Multifractal spectra $S(\tilde \Lambda,v)$
for $\eta = 2 \cdot 10^{-3}$ and $t=40$: (a) $v=0.7$; (b) $v=0.5$.
The labels 1, 2 and 3 correspond to $\delta = 10^{-7}$,
$10^{-4}$ and $10^{-2}$, respectively. The spectra have been obtained
as an average over 200,000 different realizations.

\bye